\documentclass[a4paper,12pt]{article}

\usepackage[utf8]{inputenc}
\usepackage[english]{babel}
\usepackage{graphicx}
\usepackage[top=1cm, bottom=2cm, left=2cm, right=1cm]{geometry}
\usepackage{subfigure}

\usepackage{pazheng}

\graphicspath{{figures/}}
\DeclareGraphicsExtensions{.pdf,.png,.jpg,.png}

\newcommand{\PB}[2]{\parbox[b]{#1}{#2}}

\def\apj{The Astrophysical Journal}
\def\apjl{The Astrophysical Journal Letters}
\def\apjs{The Astrophysical Journal Supplement}

\def\pasp{Pub. Astron. Soc. Pacific}

\def\aap{Astronomy and Astrophysics}

\def\aj{Astronomical Journal}
\def\mnras{MNRAS}

\begin{document}

\baselineskip 21pt

\title{\bf Study of the Dependence of the Plateau Shape for Type II Supernovae on Metallicity}

\author{\bf \hspace{-1.3cm}  \ \
Goldshtein A. A.\affilmark{1*}, Blinnikov S. I.\affilmark{2**,3,4}}

\begin{center}
\affil{
$^1${\it St. Petersburg State University, Ul'yanovskaya ul. 1, Petrodvoretz, \\ St. Petersburg, 198504 Russia}\\
$^2$ {\it Alikhanov Institute for Theoretical and Experimental Physics,  ul. Bol'shaya Cheremushkinskaya 25, Moscow, 117218 Russia} \\
$^3${\it Sternberg Astronomical Institute, Moscow State University, \\ Universitetskii pr. 13, Moscow, 119992 Russia}\\
$^4${\it Center for Fundamental and Applied Research, Dukhov All-Russia Research Institute of Automatics, ul. Sushchevskaya 22, Moscow, 127055 Russia}
       }
\end{center}
\vspace{2mm}
\sloppypar
\vspace{2mm}
\PB{.9\textwidth}{

We consider the effect of a change in the rate of flux decline in the $U$ band for type II-P supernovae (SNe IIP) as a function of metallicity $Z$. Based on this effect, we propose a new method to determine the photometric redshift from the SN IIP light curve in the $U$ band. Using the {\sc STELLA} code, we have constructed model light curves in different bands for different redshifts $z = 0.0, 0.1$, and $0.3$ as the
metallicity in the models decreases from $Z \sim 10^{-3}$ to $ \sim 10^{-6}$. The flux in the $U$ band is shown to reach a plateau at the lowest metallicities. We consider the influence of other parameters as well: the presupernova mass and the mass of radioactive nickel-56.

}

\vfill
\noindent\rule{8cm}{1pt}\\
{$^{*}$ E-mail: $<$angold107@gmail.com$>$}\\
{$^{**}$ E-mail: $<$Sergei.Blinnikov@itep.ru$>$}

\section*{INTRODUCTION}

Observations of classical type II plateau supernovae, SNe IIP, show that in the $U$, $R$, $V$, $I$, and $Z$ bands the fluxes do remain constant for two or three months, i.e., form a “plateau” that gave the name to this type. At the same time, the fluxes in the $U$ band decline steeply linearly (see, e.g., Leonard et al. 2002; Hicken et al. 2017). A less steep decline is observed in the $B$ band.

This behavior of the light curves in different bands cannot be described using the “gray” approximation in computations, which is used, for example, in the open source SNEC code (Morozova et al. 2015), where the assumption about blackbody radiation by the photosphere (with bolometric corrections) is applied to construct the light curves in different bands. The light curve constructed in this way for the $U$ band does not reproduce the observations, because the enhanced absorption by metals in the ultraviolet in the cold above-photosphere layers of the SN ejecta is required to be taken into account here.

One of the first successful computations of the effect of a linear flux decline in the $U$ band was the application of the EDDINGTON code for SNe IIP (Eastman et al. 1994). The STELLA code (Blinnikov and Sorokina 2000; Blinnikov et al. 2006), which does not use the gray approximation, but is based on multigroup radiative transfer, allows the observed effect to be described realistically.

Depending on the presupernova parameters, the
slope of the linear decline in the $U$ band can change.
Baklanov et al. (2005) noticed how the metallicity
of the presupernova envelope affects the light-curve
shape or, more specifically, the lower the metallicity,
the slower the decline. In this paper we consider in
detail the influence of this parameter with the goal
of a practical application of the gained knowledge to
supernovae with a low metallicity in their envelopes.
On the other hand, the redshift $z$ also affects the slope
of the light curve in the observed bands, because at a
sufficiently large $z$, for example, the observed $V$ band
will correspond to the $U$ band in the supernova rest
frame and there will be a linear decline instead of the
plateau in the $V$ band. Consequently, the metallicity
and the redshift can be related between each other:
if the metallicity is known from other observations,
then $z$ can be estimated. Conversely, if the redshift of
the galaxy inwhich the supernova exploded is known,
then the metallicity of its envelope can be estimated.

The redshifts of galaxies and supernovae are much
more difficult to measure than their photometric
fluxes. Therefore, methods for measuring the socalled
photometric redshifts have been proposed
long ago for galaxies (Koo 1985; Padmanabhan
et al. 2005). Attempts to determine the redshift
from supernovae are known, but they concern only
type Ia. For example, one can use the $g', r', i',$
and $z'$ bands and determine the redshift from the
fluxes at maximum light in these bands via the fitted
coefficients (Wang 2007; Wang et al. 2015). Using
the SALT2 code, Palanque-Delabrouille et al. (2018)
constructed light curves for SNe Ia in the same
bands. The boundaries for the redshift were determined
for different sets of supernova parameters.
Light curves with a given $z$ step were constructed
from approximate parameters of the light curve under
study and it was compared with them by the $\chi^2$
method.

In addition, one can use the knowledge that the
cosmological redshift affects the entire spectrum,
while absorption affects mostly the blue part of the
spectrum and compare the color indices, as was done
by Kessler et al. (2010).

Type II supernovae are used for cosmology not as
actively as SNe Ia, because their absolute luminosity
is lower than that for SNe Ia. However, since this is
the largest class of supernovae and since the power
of telescopes grows, an independent determination of
the redshifts to these objects becomes an increasingly
topical problem. In this paper we show that the effect
of a change in the slope of the light curve undoubtedly
takes place. A more quantitative analysis and lightcurve
calibrations for cosmological applications will
be performed in our succeeding publications.

\section*{MODEL CONSTRUCTION}

In the STELLA code the presupernova model is
constructed for a set of specified parameters. The
hydrostatic equilibrium equation with the assumption
about a weak dependence of the temperature on
density, $T \propto \rho^\alpha$, is used in the code. For a fully
ionized gas and homogeneous chemical composition
this hydrostatic state is close to a polytrope with
an index $1/\alpha \approx 0.3$, which satisfactorily describes
the evolutionary models, for example, from Tolstov
et al. (2016). Such an approach is used in many
papers (see, e.g., Utrobin 2007). The deviation from
the polytropic model increases in the outer layers
due to recombination and inhomogeneous chemical
composition. A point-like heavy core $0.1R_\odot$ in size
is confined at the center of the model. All of the
elements, except for $^{56}$Ni, behind the shock front are
assumed to bemixed uniformly. Since the distribution
and amount of $^{56}$Ni produced during the explosion
affect significantly the luminosity, its distribution is
specified as decreasing exponentially toward the outer
layers.

\begin{figure}[h!]
\begin{center}
\begin{minipage}[h!]{0.49\linewidth}
\includegraphics[width=1.15\linewidth]{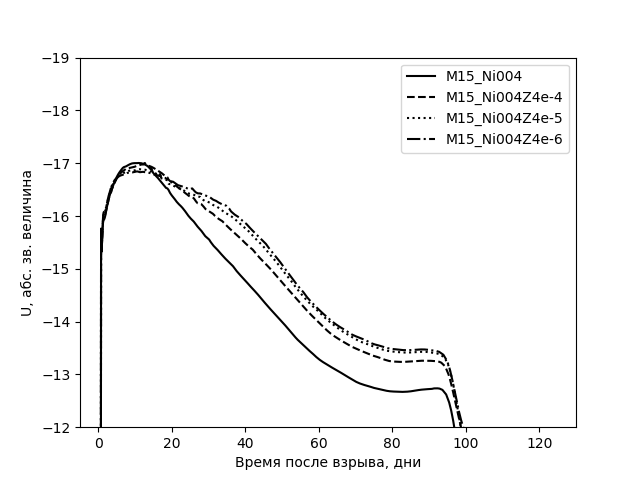}
\caption{Light curves for models 1–4 in the $U$ band. X axis: Time after explosion, days. Y axis: U, abs. magnitude} 
\label{gra:m15ni004} 
\end{minipage}
\hfill
\begin{minipage}[h!]{0.49\linewidth}
\includegraphics[width=1.15\linewidth]{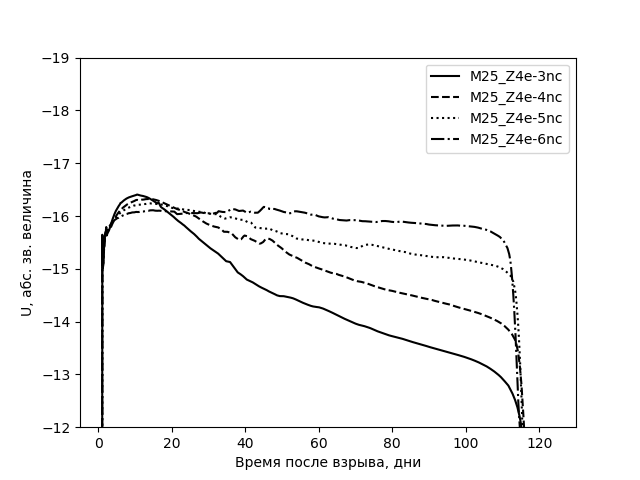}
\caption{Light curves for models 6–9 in the $U$ band.} 
\label{gra:m25ni-8nc} 
\end{minipage}
\vfill
\begin{minipage}[h!]{0.49\linewidth}
\includegraphics[width=1.15\linewidth]{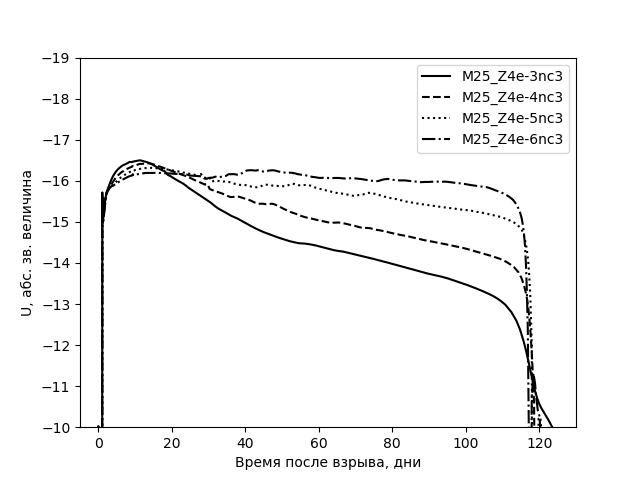}
\caption{Light curves for models 10–13 in the $U$ band.}
\label{gra:m25ni-8nc300}
\end{minipage}
\hfill
\begin{minipage}[h!]{0.49\linewidth}
\includegraphics[width=1.15\linewidth]{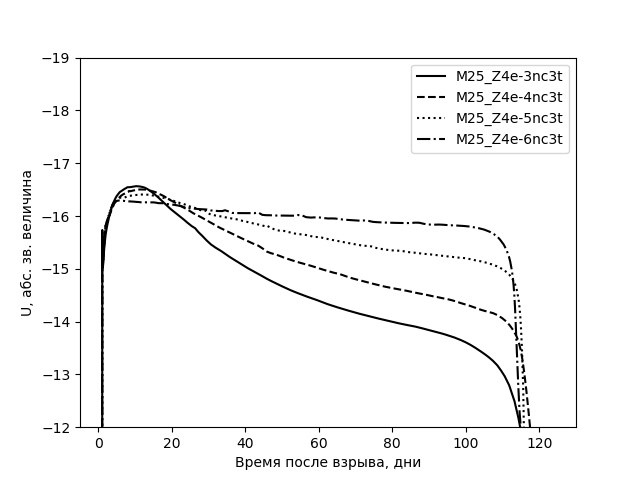}
\caption{Light curves for models 14–17 in the $U$ band.}
\label{gra:m25ni-8nc300t}
\end{minipage}
\end{center}
\end{figure}

Using the {\sc STELLA} code, we constructed several
models, they are listed in Table~\ref{tab:param}. In the first family
of models 1–4 we took standard parameters of a
supernova with different metallicities (Fig.~\ref{gra:m15ni004}). Note
that the decline becomes less sharp as the metallicity
decreases.

Then, we considered another set of parameters
with the suffix “nc” in the name (Fig.~\ref{gra:m25ni-8nc}). Apart from
an increase in the total mass and virtual zeroing of
the nickel-56 mass, the relative mass of the core with
heavy elements was artificially reduced by a factor
of 10. Thus, we can investigate the influence of
metallicity in the envelope on the light curve in pure
form. Compared to the first family, the dependence
on $Z$ in this set of models is much stronger and at
$Z = 4 \times 10^{-6}$ the plateau in the $U$ band has a very
small slope. In the next models (“nc3”) we used
the same parameters, but the number of zones was
increased for a higher accuracy. In particular, the
oscillations in the region 20–50 days for model 13
were slightly smoothed (Fig.~\ref{gra:m25ni-8nc300}). In the models with
the suffix “c3t” (see Fig.~\ref{gra:m25ni-8nc300tUV} below) the boundary of
the approximations of the inner and outer layers was
shifted outward. Thismade the light curves smoother.

In different photometric systems for the $U$ band
(Fig.~\ref{gra:4u}) the light curves look similar and the dependence
of the slope on metallicity is confirmed. In the
$R$, $I$, and $V$ bands the slope is virtually constant at
different metallicities $Z$ and this effect is imperceptible;
therefore, they are not worth considering. For
example, models 1 and 4 (Fig.~\ref{gra:m15ni004UV}) exhibit a change
in the slope at different metallicities in the $U$ band,
but in the $V$ band it remains approximately the same.
The same effect is also retained for models 14 and 17
(Fig.~\ref{gra:m25ni-8nc300tUV}). This effect can be observed in the $B$ band,
but it manifests itself not as clearly as in $U$.

\begin{figure}[h]
\begin{minipage}[h]{0.5\linewidth}
\center{\includegraphics[width=1\linewidth]{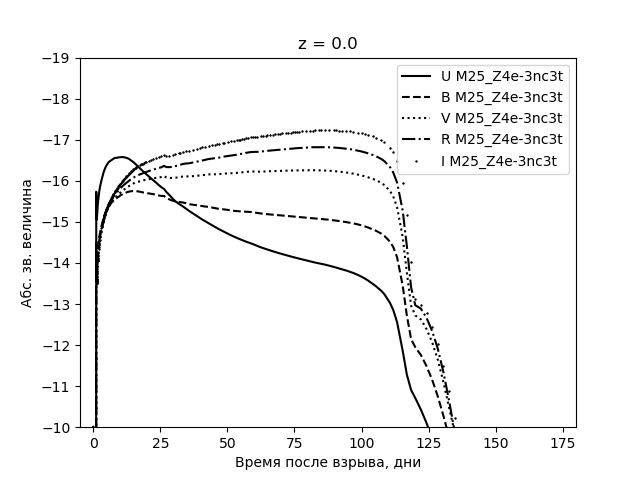}} a) \\
\end{minipage}
\hfill
\begin{minipage}[h]{0.5\linewidth}
\center{\includegraphics[width=1\linewidth]{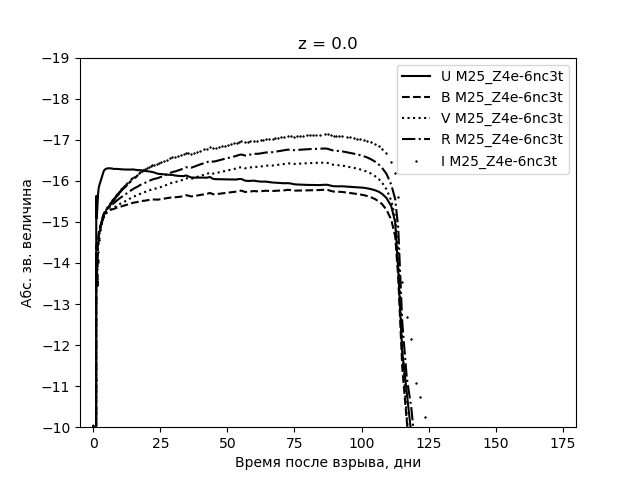}} \\b)
\end{minipage}
\vfill
\begin{minipage}[h]{0.5\linewidth}
\center{\includegraphics[width=1\linewidth]{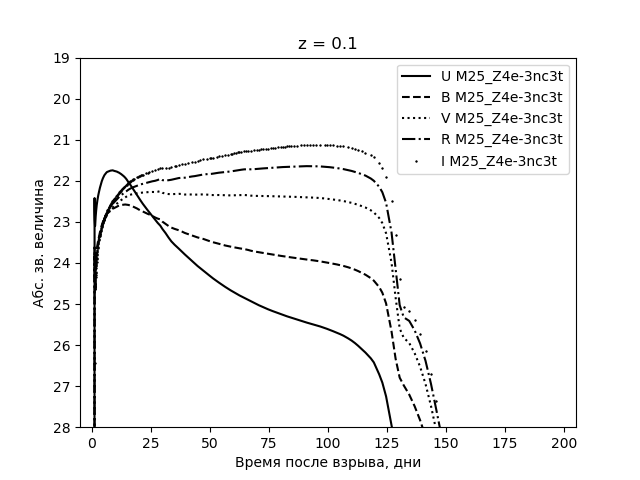}} c) \\
\end{minipage}
\hfill
\begin{minipage}[h]{0.5\linewidth}
\center{\includegraphics[width=1\linewidth]{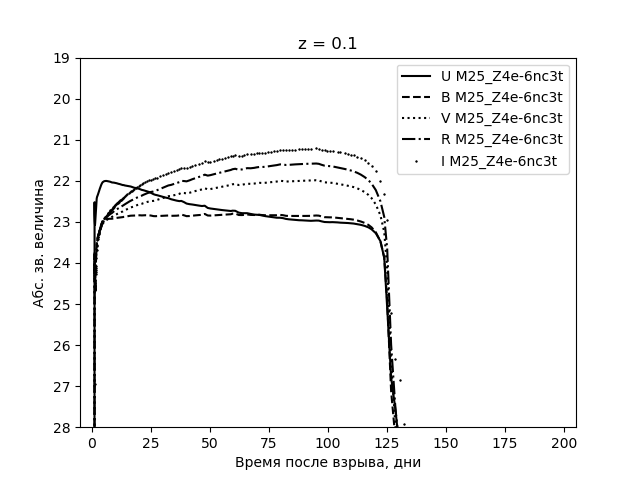}} d) \\
\end{minipage}
\vfill
\begin{minipage}[h]{0.5\linewidth}
\center{\includegraphics[width=1\linewidth]{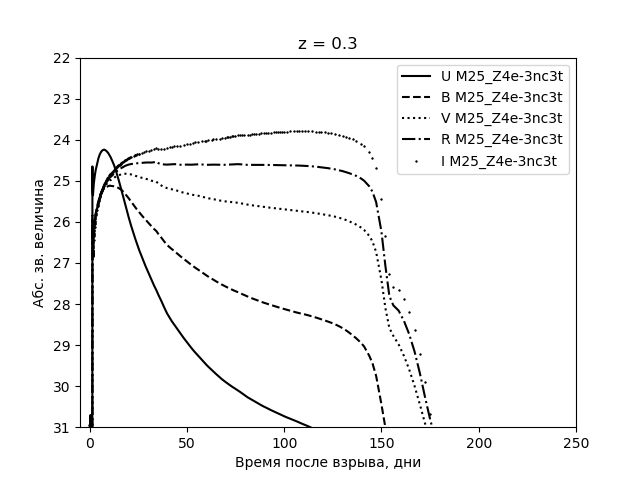}} e) \\
\end{minipage}
\hfill
\begin{minipage}[h]{0.5\linewidth}
\center{\includegraphics[width=1\linewidth]{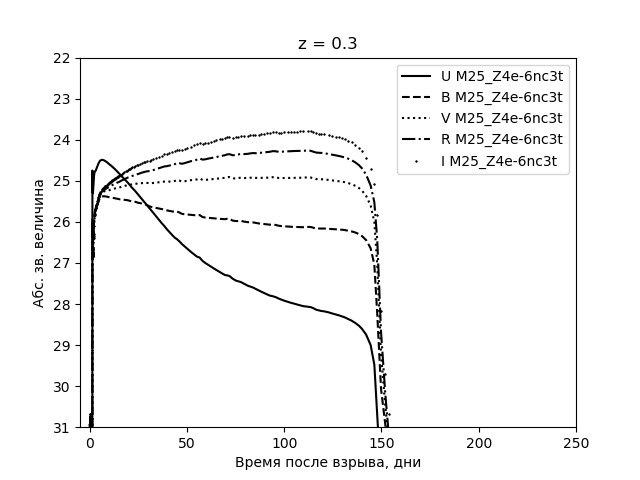}} f) \\
\end{minipage}
\caption{Light curves for extreme metallicities and different redshifts: a)~M25\_Z4e-3nc3t with z=0, b)~M25\_Z4e-6nc3t with z=0, c)~M25\_Z4e-3nc3t with z=0.1, d)~M25\_Z4e-6nc3t with z=0.1, e)~M25\_Z4e-3nc3t with z=0.3, f)~M25\_Z4e-6nc3t with z=0.3. 
X axis: Time after explosion, days. Y axis: Abs. magnitude}
\label{gra:zandZ}
\end{figure}

\begin{figure}[h]
 \begin{center}
 \includegraphics[scale=0.6]{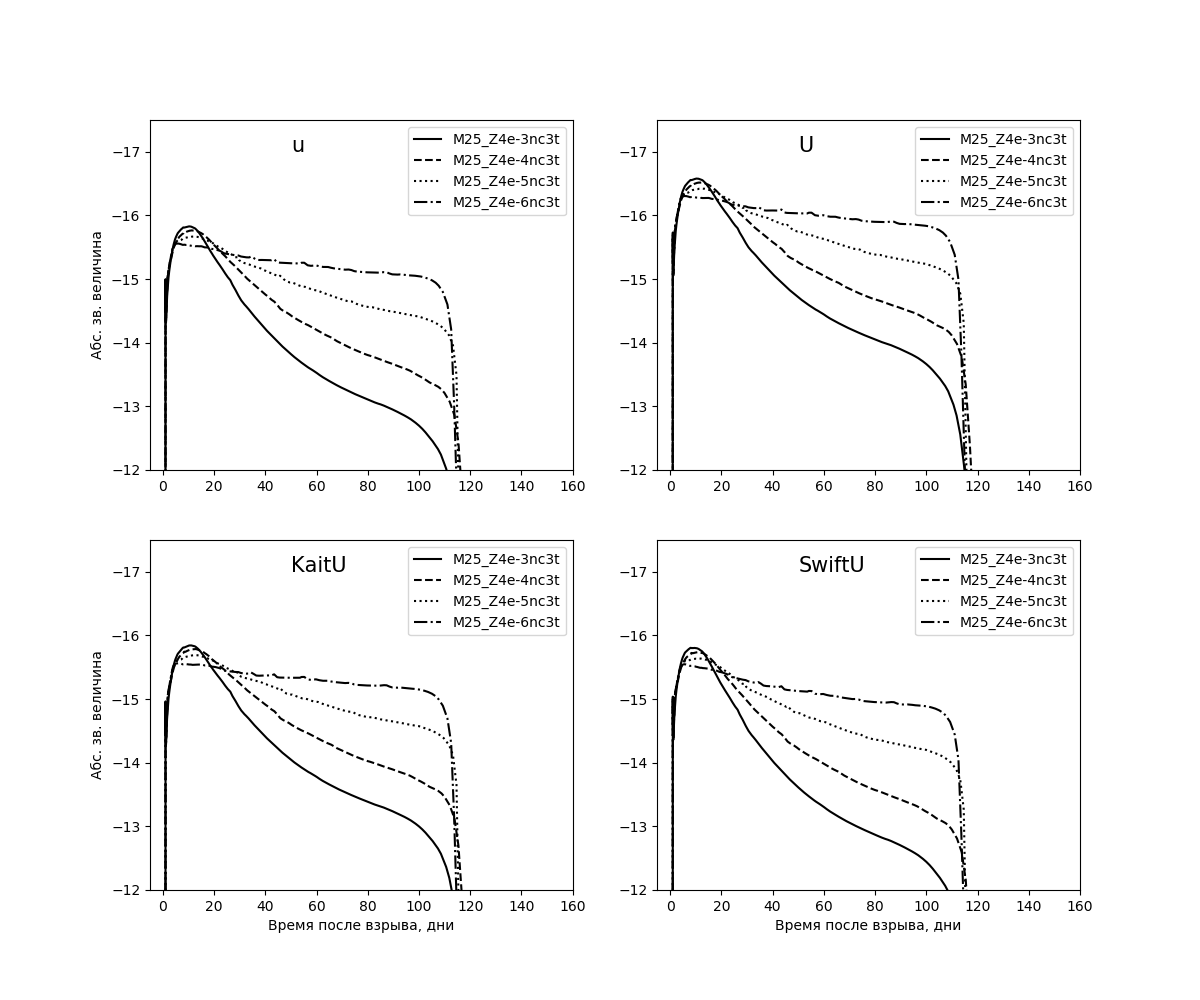}
 \caption{Light curves for models 6–9 in different photometric systems for the $U$ band. X axis: Time after explosion, days. Y axis: Abs. magnitude}
 \label{gra:4u}
 \end{center}
    \end{figure}

In the $R$, $I$, and $V$ bands this effect does not manifest
itself at different metallicities $Z$, i.e., the plateau
shape is retained (see, e.g., Fig.~\ref{gra:zandZ}). It follows from
Fig.~\ref{gra:m25ni-8nc300tUV} that a change in $Z$ affects very weakly the
plateau shape in the $V$ band. In the $B$ band the effect
is not as strong as in $U$.

It follows from the graphs that under the same
conditions a decrease in metallicity leads to an increasingly
slow decline in the $U$ band and, hence, we
can try to find an application of this dependence. Figure~\ref{gra:zandZ} shows the light curves constructed in different
bands for models 14 and 17 with different redshifts:
$z = 0, 0.1, 0.3$. As $z$ increases, the $U$ and $B$ light
curves have a faster decline due to the redshift. Note
that at $Z = 10^{-6}$ and $z = 0.3$  the shape of the light
curves is similar to that of the model with $Z = 10^{-3}$
and $z = 0$. Therefore, given the slope, we cannot unambiguously
determine the metallicity. However, this
becomes possible at a known redshift. Conversely, the
redshift $z$ can be found from the slope and metallicity.
This opens a simple way for estimating the metallicity
or redshift of metal-poor supernovae.

Tolstov et al. (2016) considered the light-curve
shape at low or zero metallicity, where the model with
zerometallicity, amass of $25M_\odot$, and $M_{\rm Ni} =0, 10^{-3}, and 10^{-1} M_\odot$ is of interest to us. In this model at
the first and second values of the nickel-56 mass the
plateau is kept at the same level, as in our M25\_Z4e-
3nc3 and M25\_Z4e-6nc3 models. The possibility to
determine the parameters from the plateau length is
discussed in the same paper, but it is often unlikely to
catch a supernova explosion before the plateau. In our
paper we consider precisely the slope and, therefore,
the time of the first measurement plays no major role.

We know yet another paper (Dessart et al. 2013),
where different metallicities were considered for
SNe IIP, but in a narrower range than that in this
paper ($Z = 0.04, 0.02, 0.008, 0.002$). In their models
the slope in the $U$ band is approximately the same and
only slightly smaller in the model with $Z = 0.002$.
At such a negligible difference it is impossible to
establish a clear dependence. Therefore, this method
cannot be applied to models with medium or high
metallicities. In addition, Dessart et al. (2013)
considered models with different kinetic energies
($E_{\rm kin} = 0.6, 1.3, 2.9$) at solar metallicity and with
identical other parameters. The slope also changes
in them and, therefore, in future it is worth checking
the influence of energy at low metallicities as well.

Recently, Potashov and Yudin (2020) have investigated
the influence of metallicity on the important
effect of time-dependence during the formation of
spectral lines in SNe IIP discovered by Utrobin and
Chugai (2002, 2005).

As is well known, for example, from Imshennik
and Nadyozhin (1988) and Utrobin (2007), the lightcurve
segment of interest to us follows the passage
of various supernova explosion stages (see Fig. 6 in
Utrobin (2007)): after $t_1$ at shock breakout adiabatic
expansion begins (from $t_1$ to the point $t_2$ at which the
decline rate slows down and the plateau begins). It
is here at the cooling wave and recombination phase
from $t_2$ to the end of the plateau $t_3$ that the light-curve
segment being studied in this paper is located.

\begin{figure}[h!]
\begin{center}
\begin{minipage}[h]{0.49\linewidth}
\includegraphics[width=1\linewidth]{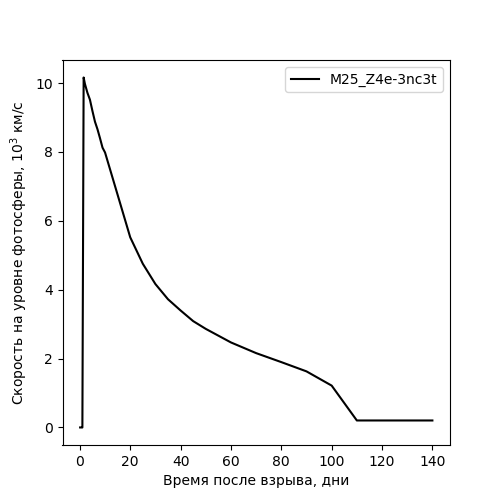}
\caption{Velocity at the photospheric level for model 14
(with a metallicity of $10^{-3}$). X axis: Time after explosion, days. Y axis: Velocity at photospheric level, $10^3$ km/s} 
\label{gra:ph3} 
\end{minipage}
\hfill
\begin{minipage}[h]{0.49\linewidth}
\includegraphics[width=1\linewidth]{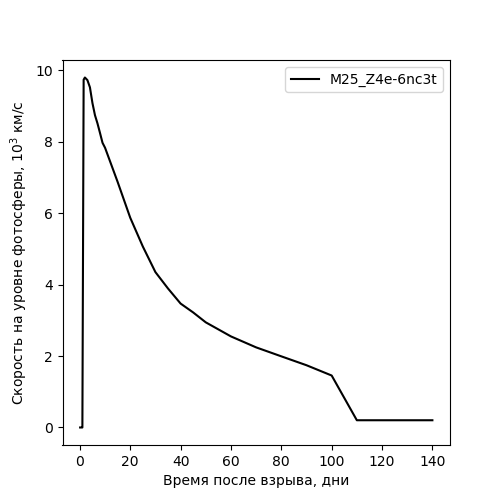}
\caption{Velocity at the photospheric level for model 17
(with a metallicity of $10^{-6}$). X axis: Time after explosion, days. Y axis: Velocity at photospheric level, $10^3$ km/s}
\label{gra:ph6}
\end{minipage}
\end{center}
\end{figure}

It follows from Figs.~\ref{gra:ph3} and ~\ref{gra:ph6} that the velocity at
the photospheric level is quite typical for SNe IIP and
a low metallicity does not affect it.

The physical causes of the dependence of the
plateau decline rate on metallicity are fairly complex
and require a careful consideration. This is planned to
be done in our subsequent paper.

\newcounter{rownum}
\setcounter{rownum}{0}
\newcommand{\Rownum}{\stepcounter{rownum}
\arabic{rownum}}

\begin{table}[!htb]
\caption{Basic parameters}
\label{tab:param}
   \centering
       \begin{tabular}{c|c|c|c|c|c}
 \hline
$No.$ & Model & Z & $M, M_\odot$ & $M_{\rm Ni}, M_\odot$ & Number of zones \\
\hline
\Rownum & M15\_Ni004 &  $4\times10^{-3}$ & 15 & 0.04 & 100\\
\Rownum & M15\_Ni004Z4e-4 & $4\times10^{-4}$ & 15 & 0.04 & 100\\
\Rownum & M15\_Ni004Z4e-5 & $4\times10^{-5}$ & 15 & 0.04 & 100\\
\Rownum & M15\_Ni004Z4e-6 & $4\times10^{-6}$ & 15 & 0.04 & 100\\
\Rownum & M15\_Ni1e-8Z4e-6 & $4\times10^{-6}$ & 15 & $10^{-8}$ & 100\\
\Rownum & M25\_Z4e-3nc & $4\times10^{-3}$ & 25 & $10^{-8}$ & 100\\
\Rownum & M25\_Z4e-4nc & $4\times10^{-4}$ & 25 & $10^{-8}$ & 100\\
\Rownum & M25\_Z4e-5nc & $4\times10^{-5}$ & 25 & $10^{-8}$ & 100\\
\Rownum & M25\_Z4e-6nc & $4\times10^{-6}$ & 25 & $10^{-8}$ & 100\\
\Rownum & M25\_Z4e-3nc3 & $4\times10^{-3}$ & 25 & $10^{-8}$ & 300\\
\Rownum & M25\_Z4e-4nc3 & $4\times10^{-4}$ & 25 & $10^{-8}$ & 300\\
\Rownum & M25\_Z4e-5nc3 & $4\times10^{-5}$ & 25 & $10^{-8}$ & 300\\
\Rownum & M25\_Z4e-6nc3 & $4\times10^{-6}$ & 25 & $10^{-8}$ & 300\\
\Rownum & M25\_Z4e-3nc3t & $4\times10^{-3}$ & 25 & $10^{-8}$ & 300\\
\Rownum & M25\_Z4e-4nc3t & $4\times10^{-4}$ & 25 & $10^{-8}$ & 300\\
\Rownum & M25\_Z4e-5nc3t & $4\times10^{-5}$ & 25 & $10^{-8}$ & 300\\
\Rownum & M25\_Z4e-6nc3t & $4\times10^{-6}$ & 25 & $10^{-8}$ & 300\\
\hline
       \end{tabular}
\end{table}

\begin{figure}[h!]
\begin{center}
\begin{minipage}[h!]{0.49\linewidth}
 \includegraphics[width=1.15\linewidth]{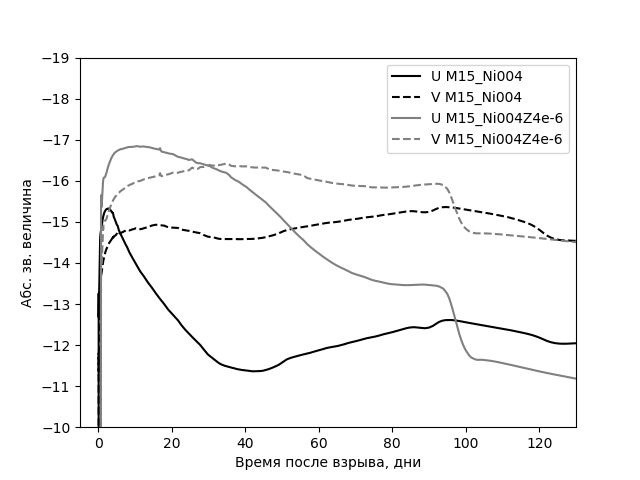}
 \caption{Light curves for models 1 and 4 in the $U$ and $V$
bands. X axis: Time after explosion, days. Y axis: Abs. magnitude} 
\label{gra:m15ni004UV} 
\end{minipage}
\hfill
\begin{minipage}[h!]{0.49\linewidth}
\includegraphics[width=1.15\linewidth]{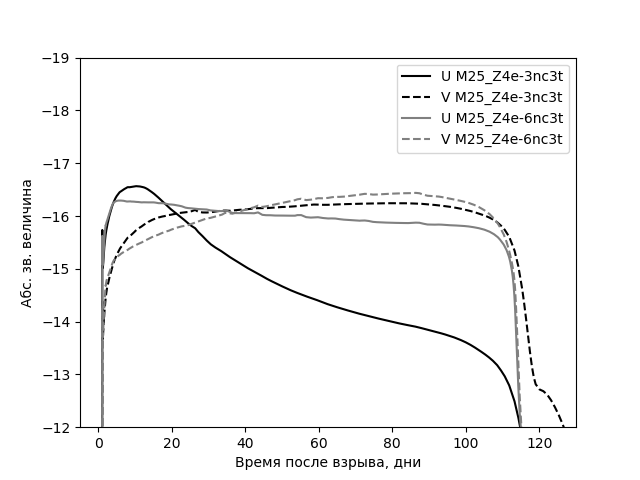}
\caption{Light curves for models 14 and 17 in the $U$ and
$V$ bands. X axis: Time after explosion, days. Y axis: Abs. magnitude}
\label{gra:m25ni-8nc300tUV}
\end{minipage}
\end{center}
\end{figure}

\newpage

\section*{CONCLUSIONS}

In this paper we considered the change in the
shape of the $U$-band light curve as a function of low
metallicity, with all other parameters being fixed. The
light curve was shown to change similarly in this
case. Therefore, by analyzing the slope after the peak,
one can find the redshift given the metallicity or the
metallicity given the redshift. This method does not
depend on the start date of observations; hence it can
be very convenient in view of the difficulty of detecting
metal-poor supernovae due to their low luminosity
(Tolstov et al. 2016). In addition, this rule holds in
different photometric systems for bands close to $U$.

A minor change of the slope in the $U$ band occurs
for stars of nearly solar metallicity. In this case, the
photometric redshift can be estimated by comparing
the slopes in the $U$, $B$, $V$, $R$, and $I$ bands. First of all,
based on long-wavelength bands, one should make
sure that the supernova does not belong to the SN IIL
class. Such supernovae observed locally, i.e., at low
redshifts, exhibit a linear decline in the $V$ band. How
can they be distinguished from distant SNe IIP, i.e.,
from the flux decline stemming from the fact that at a
high redshift $z$ the rest-frame fluxes from the $B$ and $U$
bands fall into the observed V band? For this purpose,
the fluxes in the $R$, $I$ bands and the near infrared
should be examined. In SNe IIP they will reach a
horizontal plateau, while in well-observed SNe IIL
they have the same linear decline as that in $V$ (Faran
et al. 2014; Bose et al. 2018). If the long-wavelength
fluxes reach a plateau, then the decline in the $V$ band
will be due to the redshift and $z$ can be measured after
an appropriate calibration. Since such telescopes as
LSST will discover approximately as many SNe IIP
as SNe Ia (LSST Science Collaboration 2009), it
will be possible to compare the derived photometric
redshifts of SNe IIP with the redshifts of their host
galaxies en masse. Furthermore, it should also be
kept in mind that other parameters, such as the radius,
the mass, and the explosion energy, can also
affect the light-curve shape. Therefore, this method
should be applied when the main characteristics have
been determined for sure.

Despite the fact that the values of $Z$ under consideration
are very low, stars with such a metallicity
do exist in the nearest neighborhoods. For example,
Nordlander et al. (2019) has recently discovered a
star with $Z \approx
10^{-8}$ ($[{\rm Fe/H}] = -6.2$) in our Galaxy
and, therefore, a consideration of models with such
low metallicities is not groundless. Of course, in
our neighborhoods there are no massive stars with
such low $Z$ that could explode as core-collapse supernovae,
but, at the same time, there are very many
such massive stars in the first generation and their
explosions in the nearest future will be observed with
the ground-based and space telescopes under construction.

\newpage

\section*{ACKNOWLEDGMENTS}  \label{sec:acknow}

A.A. Goldshtein thanks P.V. Baklanov and
M.Sh. Potashov for their help in preparing this
paper and M.V. Kostina for the useful discussions.
S.I. Blinnikov thanks the Russian Science Foundation
for its support of the work on the development of
the STELLA code (project no. 19-12-00229).

%

\pagebreak

\end{document}